\documentclass[aps,tightenlines,prl,twocolumn,superscriptaddress,showpacs,amsmath,amssymb,preprintnumbers,footinbib,floatfix]{revtex4}

\usepackage{graphicx}
\usepackage{dcolumn}
\usepackage{array}
\usepackage{multirow} 

\newcommand{\FiveS}{{\Upsilon(5S)}}
\newcommand{\lint}{{L_{\textrm{int}}}}

\newcommand{\sigmabb}{{\sigma_{b\bar b}^{\FiveS}}}
\newcommand{\BR}{{\mathcal B}}
\newcommand{\ebeam}{{{E_{\textrm{b}}^{\ast}}}}
\newcommand{\mbc}{{M_{\textrm{bc}}}}
\newcommand{\deltae}{{\Delta E}}

\newcommand{\tdsst}{\theta_{D_s^{\ast-}}}
\newcommand{\trho}{\theta_{\rho^+}}
\newcommand{\cdsst}{\cos\tdsst}
\newcommand{\crho}{\cos\trho}
\newcommand{\csqdsst}{\cos^2\tdsst}
\newcommand{\csqrho}{\cos^2\trho}

\newcommand{\ssqdsst}{\sin^2\tdsst}
\newcommand{\ssqrho}{\sin^2\trho}

\newcommand{\mev}{{\hbox{ MeV}}}
\newcommand{\gev}{{\hbox{ GeV}}}
\newcommand{\mmev}{{\hbox{ MeV}/c^2}}
\newcommand{\mgev}{{\hbox{ GeV}/c^2}}

\newcommand{\bs}{{B_s^0}}
\newcommand{\bsst}{{B_s^{\ast}}}
\newcommand{\barbsst}{{{\bar B_s}^{\ast}}}
\newcommand{\bsST}{{B_s^{(\ast)}}}
\newcommand{\bsSTbsST}{{\bsST{\bar B_s}^{(\ast)}}}
\newcommand{\KS}{{K_S^0}}
\newcommand{\ds}{{D_s^-}}
\newcommand{\dsst}{{D_s^{\ast-}}}

\newcommand{\bsdspi}{{\bs\to D_s^-\pi^+}}
\newcommand{\bsdsrho}{{\bs\to D_s^-\rho^+}}
\newcommand{\bsdsstrho}{{\bs\to D_s^{\ast-}\rho^+}}
\newcommand{\bsdsSTrho}{{\bs\to D_s^{(\ast)-}\rho^+}}
\newcommand{\bsdsstpi}{{\bs\to D_s^{\ast-}\pi^+}}
\newcommand{\bsdsSTpi}{{\bs\to D_s^{(\ast)-}\pi^+}}

\newcommand{\bfbsdsstpi}{(2.4^{+0.5}_{-0.4}({\rm stat.})\pm0.3({\rm syst.})\pm0.4(f_s))\times10^{-3}}
\newcommand{\bfbsdsrho}{(8.5^{+1.3}_{-1.2}({\rm stat.})\pm1.1({\rm syst.})\pm1.3(f_s))\times10^{-3}}
\newcommand{\bfbsdsstrho}{(11.8^{+2.2}_{-2.0}({\rm stat.})\pm1.7({\rm syst.})\pm1.8(f_s))\times10^{-3}}
\newcommand{\fl}{1.05^{+0.08}_{-0.10}{\rm{(stat.)}{}^{+0.03}_{-0.04}{\rm(syst.)}}}

\begin{document}
  \date{\today}  
  
  \preprint{
    \vbox{
      \hbox{KEK Preprint 2010-5}
      \hbox{BELLE Preprint 2010-7}
    }
  }

  \title{Observation of \boldmath{$\bsdsstpi$}, \boldmath{$\bsdsSTrho$} Decays\\
    and Measurement of $\bsdsstrho$ Polarization}

  \affiliation{Budker Institute of Nuclear Physics, Novosibirsk}
  \affiliation{Faculty of Mathematics and Physics, Charles University, Prague}
  \affiliation{University of Cincinnati, Cincinnati, Ohio 45221}
  \affiliation{Justus-Liebig-Universit\"at Gie\ss{}en, Gie\ss{}en}
  \affiliation{The Graduate University for Advanced Studies, Hayama}
  \affiliation{Hanyang University, Seoul}
  \affiliation{University of Hawaii, Honolulu, Hawaii 96822}
  \affiliation{High Energy Accelerator Research Organization (KEK), Tsukuba}
  \affiliation{Institute of High Energy Physics, Chinese Academy of Sciences, Beijing}
  \affiliation{Institute of High Energy Physics, Vienna}
  \affiliation{Institute of High Energy Physics, Protvino}
  \affiliation{Institute for Theoretical and Experimental Physics, Moscow}
  \affiliation{J. Stefan Institute, Ljubljana}
  \affiliation{Kanagawa University, Yokohama}
  \affiliation{Institut f\"ur Experimentelle Kernphysik, Karlsruher Institut f\"ur Technologie, Karlsruhe}
  \affiliation{Korea Institute of Science and Technology Information, Daejeon}
  \affiliation{Korea University, Seoul}
  \affiliation{Kyungpook National University, Taegu}
  \affiliation{\'Ecole Polytechnique F\'ed\'erale de Lausanne (EPFL), Lausanne}
  \affiliation{Faculty of Mathematics and Physics, University of Ljubljana, Ljubljana}
  \affiliation{University of Maribor, Maribor}
  \affiliation{Max-Planck-Institut f\"ur Physik, M\"unchen}
  \affiliation{University of Melbourne, School of Physics, Victoria 3010}
  \affiliation{Nagoya University, Nagoya}
  \affiliation{Nara Women's University, Nara}
  \affiliation{National Central University, Chung-li}
  \affiliation{National United University, Miao Li}
  \affiliation{Department of Physics, National Taiwan University, Taipei}
  \affiliation{H. Niewodniczanski Institute of Nuclear Physics, Krakow}
  \affiliation{Nippon Dental University, Niigata}
  \affiliation{Niigata University, Niigata}
  \affiliation{University of Nova Gorica, Nova Gorica}
  \affiliation{Novosibirsk State University, Novosibirsk}
  \affiliation{Osaka City University, Osaka}
  \affiliation{Panjab University, Chandigarh}
  \affiliation{University of Science and Technology of China, Hefei}
  \affiliation{Seoul National University, Seoul}
  \affiliation{Sungkyunkwan University, Suwon}
  \affiliation{School of Physics, University of Sydney, NSW 2006}
  \affiliation{Tata Institute of Fundamental Research, Mumbai}
  \affiliation{Tohoku Gakuin University, Tagajo}
  \affiliation{Tohoku University, Sendai}
  \affiliation{Department of Physics, University of Tokyo, Tokyo}
  \affiliation{Tokyo Metropolitan University, Tokyo}
  \affiliation{Tokyo University of Agriculture and Technology, Tokyo}
  \affiliation{IPNAS, Virginia Polytechnic Institute and State University, Blacksburg, Virginia 24061}
  \affiliation{Yonsei University, Seoul}
  
  \author{R.~Louvot}\affiliation{\'Ecole Polytechnique F\'ed\'erale de Lausanne (EPFL), Lausanne} % Lausanne
  \author{O.~Schneider}\affiliation{\'Ecole Polytechnique F\'ed\'erale de Lausanne (EPFL), Lausanne} % Lausanne
  \author{T.~Aushev}\affiliation{\'Ecole Polytechnique F\'ed\'erale de Lausanne (EPFL), Lausanne}\affiliation{Institute for Theoretical and Experimental Physics, Moscow} % ITEP
  
  \author{K.~Arinstein}\affiliation{Budker Institute of Nuclear Physics, Novosibirsk}\affiliation{Novosibirsk State University, Novosibirsk} % BINP
  \author{A.~M.~Bakich}\affiliation{School of Physics, University of Sydney, NSW 2006} % Sydney
  \author{V.~Balagura}\affiliation{Institute for Theoretical and Experimental Physics, Moscow} % ITEP
  \author{E.~Barberio}\affiliation{University of Melbourne, School of Physics, Victoria 3010} % Melbourne
  \author{A.~Bay}\affiliation{\'Ecole Polytechnique F\'ed\'erale de Lausanne (EPFL), Lausanne} % Lausanne
  \author{K.~Belous}\affiliation{Institute of High Energy Physics, Protvino} % Protvino
  \author{M.~Bischofberger}\affiliation{Nara Women's University, Nara} % Nara
% ...
% BINP

  \collaboration{Belle Collaboration}
  
  \noaffiliation
  
  \begin{abstract}
    First observations of the $\bsdsstpi$, $\bsdsrho$ and $\bsdsstrho$ decays are reported
    together with measurements of their branching fractions: $\BR(\bsdsstpi)=\bfbsdsstpi$, $\BR(\bsdsrho)=\bfbsdsrho$ and $\BR(\bsdsstrho)=\bfbsdsstrho$ ($f_s=N_{\bsSTbsST}/N_{b\bar b}$).
    From helicity-angle distributions,
    we measured the longitudinal polarization fraction in $\bsdsstrho$ decays to be $f_L(\bsdsstrho)=\fl$.
    These results are based on a 23.6 fb$^{-1}$ data sample
    collected at the $\Upsilon(5S)$ resonance
    with the Belle detector at the KEKB $e^+e^-$ collider.
  \end{abstract}
  
  
  \maketitle
      {\renewcommand{\thefootnote}{\fnsymbol{footnote}}}
      \setcounter{footnote}{0}
      
      The measurement of exclusive $\bs\to D_s^{(\ast)-}h^+$ \cite{chargeconj}
      ($h^+=\pi^+$ or $\rho^+$) decays is an important milestone in the study
      of the poorly understood decay processes of the $\bs$ meson.
      In Refs.~\cite{PRL_98_052001,PRD_76_012002,PRL_100_121801,PRL_102_021801}
      Belle confirmed the large potential of $B$ factories for $\bs$ investigations
      due to the low multiplicities of charged and neutral particles and high reconstruction
      efficiencies.
      We have now observed three new exclusive $\bs$ modes
      with relatively large branching fractions and neutral particles such as photons or $\pi^0$'s 
      in their final states.
      The leading amplitude for the four $\bsdsSTpi$ and $\bsdsSTrho$ modes
      is a $b\to c$ tree diagram of order $\lambda^2$
      (in the Wolfenstein parameterization \cite{PRL_51_1945}
      of the CKM quark-mixing matrix \cite{PRL_10_531}) with a spectator $s$ quark.
      The study of $\bs$ decays provides useful tests of the
      heavy-quark theories that predict, based on an $SU(3)$ symmetry,
      similarities between $\bs$-meson decay modes
      and their corresponding $B^0$-meson counterparts.
      These include the unitarized quark model \cite{PRL_53_878},
      the heavy quark effective theory (HQET)
      \cite{PRD_42_3732,NPB_389_534,PLB_318_549,PLB_259_359},
      and a more recent approach based on chiral symmetry \cite{PRD_68_054024}.
      Our $\bs$ branching fraction results can be used
      to normalize measurements of $\bs$ decays made at
      hadron collider experiments, where the number
      of $\bs$ mesons produced has a substantial systematic
      uncertainty.

      The decay $\bs\to D_s^{\ast-}h^+$ is mediated
      by the same tree diagram as $B^0\to D^{\ast-}h^+$,
      but with a spectator $s$ quark.
      The contribution of the strongly suppressed $W$-exchange diagram is
      expected to be negligibly small.
      Moreover, the helicity amplitudes in $B\to VV$ decays can be used
      to test the factorization hypothesis \cite{PLB_89_105,PLB_259_359}.
      The relative strengths of the longitudinal and transverse states can be
      measured with an angular analysis of the decay products.
      In the helicity basis, the expected $\bsdsstrho$ differential decay width is
      \begin{eqnarray}\frac{{\rm d}^2\Gamma(\bsdsstrho)}{{\rm d}\cdsst {\rm d}\crho}&\propto& 4f_L\ssqdsst\csqrho+\\
	&&(1-f_L)(1+\csqdsst)\ssqrho\,,\nonumber\end{eqnarray}
      where $f_L=|H_0|^2/\sum_{\lambda}|H_{\lambda}|^2$ is the longitudinal polarization fraction, 
      $H_{\lambda}$ ($\lambda=\pm1,0$) are the helicity amplitudes, and 
      $\tdsst$ ($\trho$)  is the helicity angle of the $D_s^{\ast-}$ ($\rho^+$)
      defined as the supplement of the angle between the $\bs$ and the $\ds$ ($\pi^+$) momenta
      in the $\dsst$ ($\rho^+$) frame.

      Here
      we report measurements performed with fully reconstructed 
      $\bsdsstpi$, $\bsdsrho$ and $\bsdsstrho$ decays
      in a data set corresponding to an integrated luminosity of
      $\lint=(23.6\pm0.3)$~fb$^{-1}$ collected with the Belle detector at
      the KEKB asymmetric-energy (3.6\gev~on 8.2\gev) $e^+e^-$ collider \cite{NIMA_499_1}
      operated at the $\Upsilon(5S)$ resonance
      ($\sqrt s=10867.0\pm1.0\mev$ \cite{PRL_102_021801}).
      The total $b\bar b$ cross section at the $\FiveS$ energy has been measured to be
      $\sigmabb=(0.302\pm0.014)$~nb \cite{PRL_98_052001,PRD_75_012002}.
      Three $\bs$ production modes are kinematically allowed at the $\FiveS$:
      $\bsst\barbsst$, $\bsst\bar\bs+\bs\barbsst$, and $\bs\bar\bs$.
      The $\bsst$ decays to $\bs$, emitting a photon with  energy  $E_{\gamma}\sim50\mev$.
      The fraction of $b\bar b$ events containing a $\bsSTbsST$ pair has been measured to be
      $f_s=N_{\bsSTbsST}/N_{b\bar b}=(19.3\pm2.9)\%$ \cite{PLB_667_1}.
      The fraction of $\bsSTbsST$ events containing a $\bsst\barbsst$ pair is predominant
      and has been measured with $\bsdspi$ events to be 
      $f_{\bsst\barbsst}=(90.1^{+3.8}_{-4.0}\pm0.2)\%$ \cite{PRL_102_021801}.
      The number of $\bs$ mesons produced in the dominant $\bsst\barbsst$ production mode
      is thus
      $N_{\bs}=2\times\lint\times\sigmabb\times f_s\times f_{\bsst\barbsst}=(2.48\pm0.41)\times10^6$.
   
      The Belle detector is a large-solid-angle magnetic
      spectrometer that consists of a silicon vertex detector,
      a central drift chamber (CDC), an array of
      aerogel threshold Cherenkov counters (ACC),
      a barrel-like arrangement of time-of-flight
      scintillation counters (TOF), and an electromagnetic calorimeter
      comprised of CsI(Tl) crystals (ECL) located inside
      a superconducting solenoid coil that provides a 1.5~T
      magnetic field. An iron flux-return located outside of 
      the coil is instrumented to detect $K_L^0$ and to identify 
      muons. The detector is described in detail elsewhere~\cite{NIMA_479_117}.

      Reconstructed charged tracks are required to have a maximum impact parameter with respect
      to the nominal interaction point of 0.5~cm in the radial direction and 3~cm
      in the beam-axis direction.
      A likelihood ratio $\mathcal R_{K/\pi}=\mathcal{L}_K/(\mathcal L_{\pi}+\mathcal L_K)$
      is constructed using
      ACC, TOF and CDC (ionization energy loss) measurements.
      A track is identified as a charged pion if $\mathcal R_{K/\pi}<0.6$
      or as a charged kaon otherwise.
      With this selection, 
      the momentum-averaged identification efficiency for pions (kaons) is about $91\%$ ($86\%$),
      while the momentum-averaged rate of kaons (pions) identified as pions (kaons) 
      is about $9\%$ ($14\%$). 
      
      Photons are reconstructed using ECL energy clusters
      within the polar angle acceptance $17^{\circ}$ to $150^{\circ}$
      that are not associated with a charged track and that have an energy deposit
      larger than $50\mev$.
      A photon candidate is retained only if the ratio of the energy deposited
      in the array of the central $3\times3$ cells is
      more than 85\% of that in the array of $5\times5$ cells.
      Neutral pions are reconstructed via the $\pi^0\to\gamma\gamma$ decay with photon pairs
      having an invariant mass within $\pm13\mmev$
      of the $\pi^0$ mass.
      A mass-constrained fit is then applied to the $\pi^0$ candidates.
      
      Neutral kaons are reconstructed via the decay $\KS\to\pi^+\pi^-$
      with no $\mathcal R_{K/\pi}$ requirements
      for the two charged pions.
      The $\KS$ candidates are required to have an invariant mass within $\pm7.5\mmev$
      of the $\KS$ mass.
      Requirements are applied on the $\KS$ vertex displacement from the interaction point (IP)
      and on the difference between the $\KS$ flight directions
      obtained from the $\KS$ momentum and from the decay vertex and IP.
      The criteria are described in detail elsewhere \cite{phd_ffang}.
      The $K^{\ast0}$ ($\phi$, $\rho^+$) candidates are reconstructed
      via the decay $K^{\ast0}\to K^+\pi^-$ ($\phi\to K^+K^-$, $\rho^+\to\pi^+\pi^0$)
      with an invariant mass
      within $\pm50\mmev$ ($\pm12\mmev$, $\pm100\mmev$) of their nominal values.
      Candidates for $\ds$ are reconstructed in the three modes $\ds\to\phi\pi^-$,
      $\ds\to K^{\ast0}K^-$,
      and $\ds\to\KS K^-$ and are required to have a mass within $\pm10\mmev$
      of the $\ds$  mass. 
      The $\dsst$ candidates are reconstructed via the decay $\dsst\to\ds\gamma$
      by adding a photon candidate to a $\ds$ candidate.
      The $\ds\gamma$ pair is required to have a mass difference $m(\ds\gamma)-m(\ds)$
      within $\pm13\mmev$ of the $\dsst-\ds$ mass difference. 
      All mass values are those reported in Ref.~\cite{PLB_667_1},
      and the applied mass windows correspond to $\pm(3-4)\sigma$ around these values;
      the mass resolution, $\sigma$, is obtained from MC signal simulations.
       
      The $\bsdsstpi$ and $\bsdsrho$ candidates are reconstructed using two variables:
      the beam-energy-constrained mass of the $\bs$ candidate
      $\mbc=\sqrt{\ebeam^2-\vec p_{\bs}^{\ast2}}$, and the energy difference
      $\deltae=E_{\bs}^{\ast}-\ebeam$,
      where $(E_{\bs}^{\ast},\vec p_{\bs}^{\ast})$ is the four-momentum of the $\bs$ candidate
      and $\ebeam$ is the beam energy, both expressed in the center-of-mass frame.
      The two angles $\tdsst$ and $\trho$ are used as additional observables
      for the $\bsdsstrho$ candidate.
      We select candidates with $\mbc>5.3\mgev$ and $-0.3\gev<\deltae<0.4\gev$.

      Further selection criteria are developed using Monte Carlo (MC) samples
      based on the EvtGen \cite{NIMA_462_152} event generator
      and the GEANT \cite{geant} full-detector simulation.
      The most significant source of background is continuum processes,
      $e^+e^-\to q\bar q$ ($q=u,d,s,c$).
      In addition, peaking backgrounds can arise from specific $\bs$ decays.
      Using a MC sample of $e^+e^-\to\bsSTbsST$ events corresponding
      to three times the integrated luminosity,
      we find that $\bsdspi$ and $\bsdsrho$ events make a significant contribution
      to the background in the $\bsdsstpi$ analysis.
      However, they are well separated from the signal in the $\deltae$ distribution.
      If a $\bsdspi$ decay is combined with an extra photon,
      the energy is larger than the signal;
      the four charged tracks of a $\bsdsrho$ event can be selected
      with an additional photon giving a
      $\bs$ candidate with a smaller energy.
      Similarly, $\bsdsstrho$ decays give a significant contribution
      to the $\bsdsrho$ analysis at lower energies. 
      For the $\bsdsstrho$ analysis, there is no significant peaking background.
      MC studies show that, for the three modes, all the other background sources
      (mainly $B^0$ and $B^+$ events) are smooth and small enough to be well
      described by the same shape that is used for the continuum.      
      The contribution of non-resonant $\bs\to D_s^{(\ast)-}\pi^+\pi^0$ decays
      is studied by relaxing the $(\pi^+\pi^0)$ mass ($M_{\pi\pi}$) requirement
      and doing a two-dimensional fit in $\mbc$ and $\deltae$ (see below).
      The signal $M_{\pi\pi}$ distribution is then obtained using the $_{\rm s}$Plot method \cite{NIMA_555_356}.
      The resulting $M_{\pi\pi}$
      spectrum shows no indication of $\bs\to D_s^{(\ast)-}\pi^+\pi^0$ decays
      (consistent with results for $B^0\to D^{(\ast)+}\pi^0\pi^-$ \cite{PRD_50_43}),
      and we neglect this component in our fit.
      
      To improve signal significance, criteria for each of the three $\bs$ modes
      are chosen to maximize 
      $N_{\textrm{sig}}/\sqrt{N_{\textrm{sig}}+N_{\textrm{bkg}}^{q\bar q}+N_{\textrm{bkg}}^{\textrm{peak.}}}$,
      evaluated in the $\pm2.5\sigma$ $\bsst\barbsst$ signal region in the $(\mbc,\deltae)$ plane.
      The expected continuum background, $N_{\textrm{bkg}}^{q\bar q}$,
      is estimated using MC-generated continuum events corresponding to three times the data.
      The expected signal, $N_{\textrm{sig}}$, and peaking background,
      $N_{\textrm{bkg}}^{\textrm{peak}}$,
      are obtained assuming $\BR(\bsdspi)=\BR(\bsdsstpi)=3.3\!\times\!10^{-3}$ \cite{PLB_667_1}
      and $\BR(\bsdsrho)=\BR(\bsdsstrho)=7.0\!\times\!10^{-3}$ \cite{PRD_42_3732}.
      The efficiencies of exclusive $\bs$ decays are determined using MC simulations.
           
      To suppress the continuum background,
      we use the ratio of the second and zeroth Fox-Wolfram moments \cite{PRL_41_1581}, $R_2$.
      This variable has a broad distribution between zero
      and one for jet-like continuum events and 
      is concentrated in the range below $0.5$ for the more spherical signal events.
      This property allows an efficient continuum reduction
      with a low systematic uncertainty ($\sim2\%$).
      Candidates for $\bsdsstpi$ ($\bsdsrho$ and $\bsdsstrho$)
      are required to have $R_2<0.5$ ($<0.35$).
      This selection rejects $40\%$ ($69\%$, $64\%$) of the background while
      retaining $93\%$  ($82\%$, $86\%$) of the $\bsdsstpi$ ($\bsdsrho$, $\bsdsstrho$) signal.
      
      After the event selection described above, about
      15\%, 15\%, and 28\% of 
      $\dsst\pi^+$, $\ds\rho^+$ and $\dsst\rho^+$ candidate events, respectively, 
      have multiple candidates. We select one
      candidate per event according to the following
      criteria.
      The $D_s^+$ with the mass closest to the nominal value is preferred.
      The $D_s^{\ast+}$ formed with the preferred $D_s^+$ and
      with the mass difference $m(D_s^{\ast})-m(D_s)$ closest to the nominal value is preferred.
      The $\bsdsstpi$  candidate with the preferred $D_s^{\ast-}$
      and the $\pi^+$ with the best $\mathcal R_{K/\pi}$ is retained.
      The preferred $\rho^+$ is the one with the $\pi^0$ mass
      (before the mass-constrained fit) closest to the nominal value
      and the $\pi^+$ with the best $\mathcal R_{K/\pi}$.
      The $\bsdsrho$ ($\bsdsstrho$) candidate with the preferred $D_s^-$ ($D_s^{\ast-}$)
      and the preferred $\rho^+$ is retained.
      After this selection, in MC signal simulations, 76\%, 68\% and 51\% (64\%)
      of the selected $\bsdsstpi$, $\bsdsrho$
      and longitudinally (transversally) polarized $\bsdsstrho$ candidates
      are correctly reconstructed.
      
      The $\bsdsstpi$ and $\bsdsrho$ signals are extracted from
      a two-dimensional unbinned extended maximum likelihood fit \cite{NIMA_297_496}
      in $\mbc$ and $\deltae$.
      The three decays of the $\Upsilon(5S)$
      ($\bsst\barbsst$, $\bsst\bar\bs+\bs\barbsst$ and $\bs\bar\bs$) are considered.
      Each signal probability density function (PDF) is described
      with sums of Gaussian or so-called
      ``Novosibirsk functions'' \cite{novofunc};
      the latter function is used to describe the distribution
      if it is asymmetrical around its central value.
      Each signal PDF is composed of two components with their respective proportions fixed,
      representing the correctly and the incorrectly reconstructed candidates.
      In a simulated signal event, a candidate is correctly (incorrectly) reconstructed
      when the selected decay products do (do not) match the true combination.
      The fractions of correctly reconstructed candidates are fixed from MC samples
      and their uncertainties are included in the systematic error.
      The $\mbc$ and $\deltae$ resolutions for $\bsdsstpi$ ($\bsdsrho$ and $\bsdsstrho$)
      are calibrated by a multiplying factor measured with the $\bsdspi$ \cite{PRL_102_021801}
      ($B^0\to D^{\ast-}\rho^+$) signal.
      The mean values of $\mbc$ and $\deltae$ for the three $\bs$ production modes (6 parameters)
      are related to two floating parameters
      corresponding to the $\bs$ and $\bsst$ meson masses \cite{masses}.
      The peaking background PDFs are analytically defined and fixed from specific MC samples.
      The continuum (together with possible $B^+$ and $B^0$ background) is modeled
      with an ARGUS function \cite{PLB_185_218} for $\mbc$ and a linear function for $\deltae$.
      The endpoint of the ARGUS function is fixed to the beam energy,
      while the two other parameters are left free.
      All the yields can float.

      For the $\bsdsstrho$ candidates, we perform a four-dimensional fit 
      using the two observables $\cdsst$ and $\crho$ in addition to $\mbc$ and $\deltae$.
      Only the main $\bs$ production mode is considered ($\bsst\barbsst$),
      and three components are used in the likelihood:
      the transverse and longitudinal signals, and the background.
      We define the PDF for $\mbc$ and $\deltae$ in the same way as described above,
      while the angular distributions are analytically
      described with polynomials of order up to five.
      The shape parameters are floated for the background PDF
      but are fixed for the two signal PDFs. 
      
\begin{figure}
	\centering
	\includegraphics[width=\linewidth]{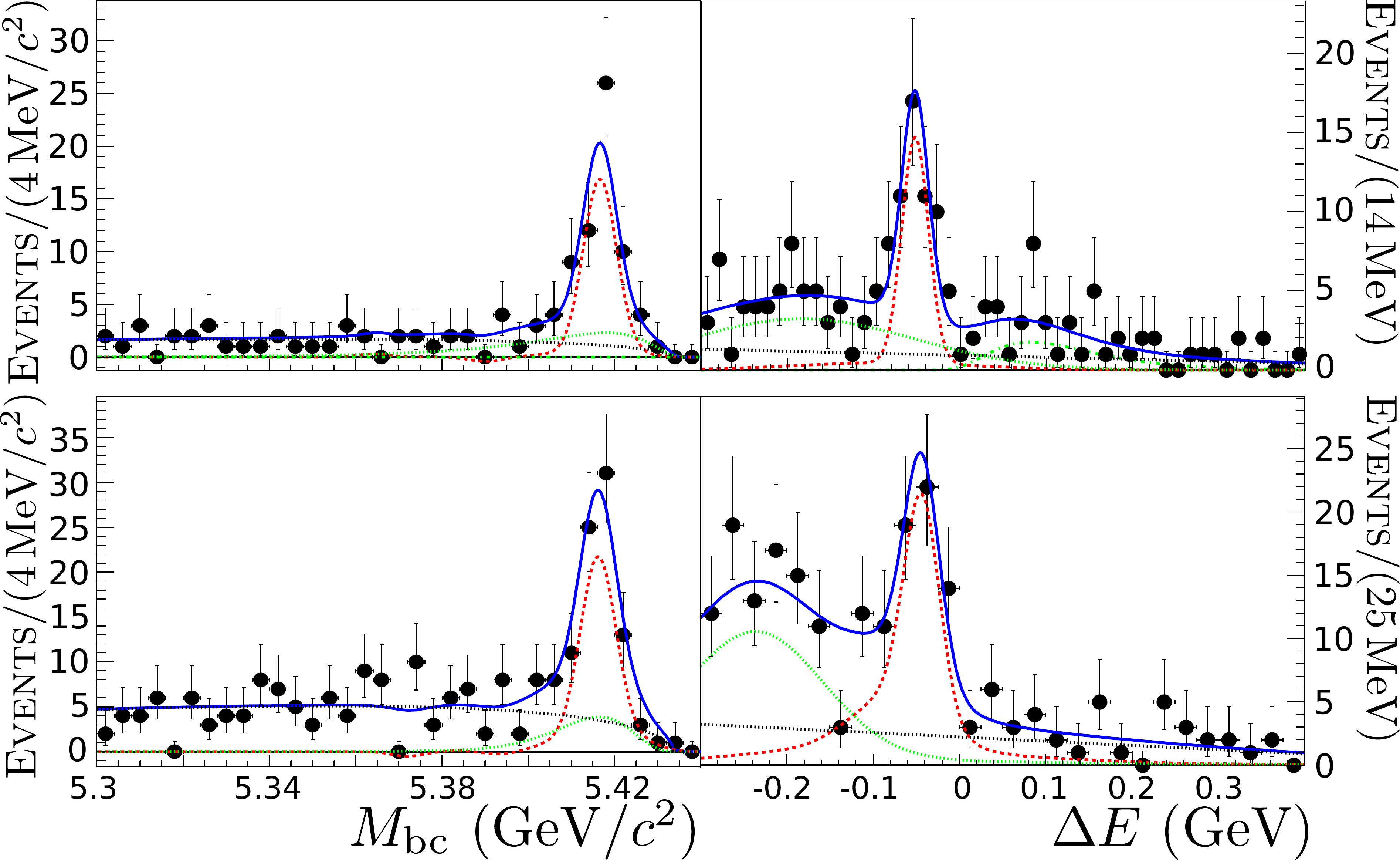}
	\caption{\label{fig2}Left (right): $\mbc$ ($\deltae$) distributions
	  for the $\bsdsstpi$ (top) and $\bsdsrho$ (bottom) candidates
	  with $\deltae$ ($\mbc$) restricted to the $\pm2.5\sigma$ $\bsst\barbsst$ signal region.
	  The blue solid curve is the total PDF, 
	  while the green (black) dotted curve is the peaking (continuum)
          background and the red dashed curve is the signal.}
      \end{figure}

\begin{figure}
	\centering
	\includegraphics[width=\linewidth]{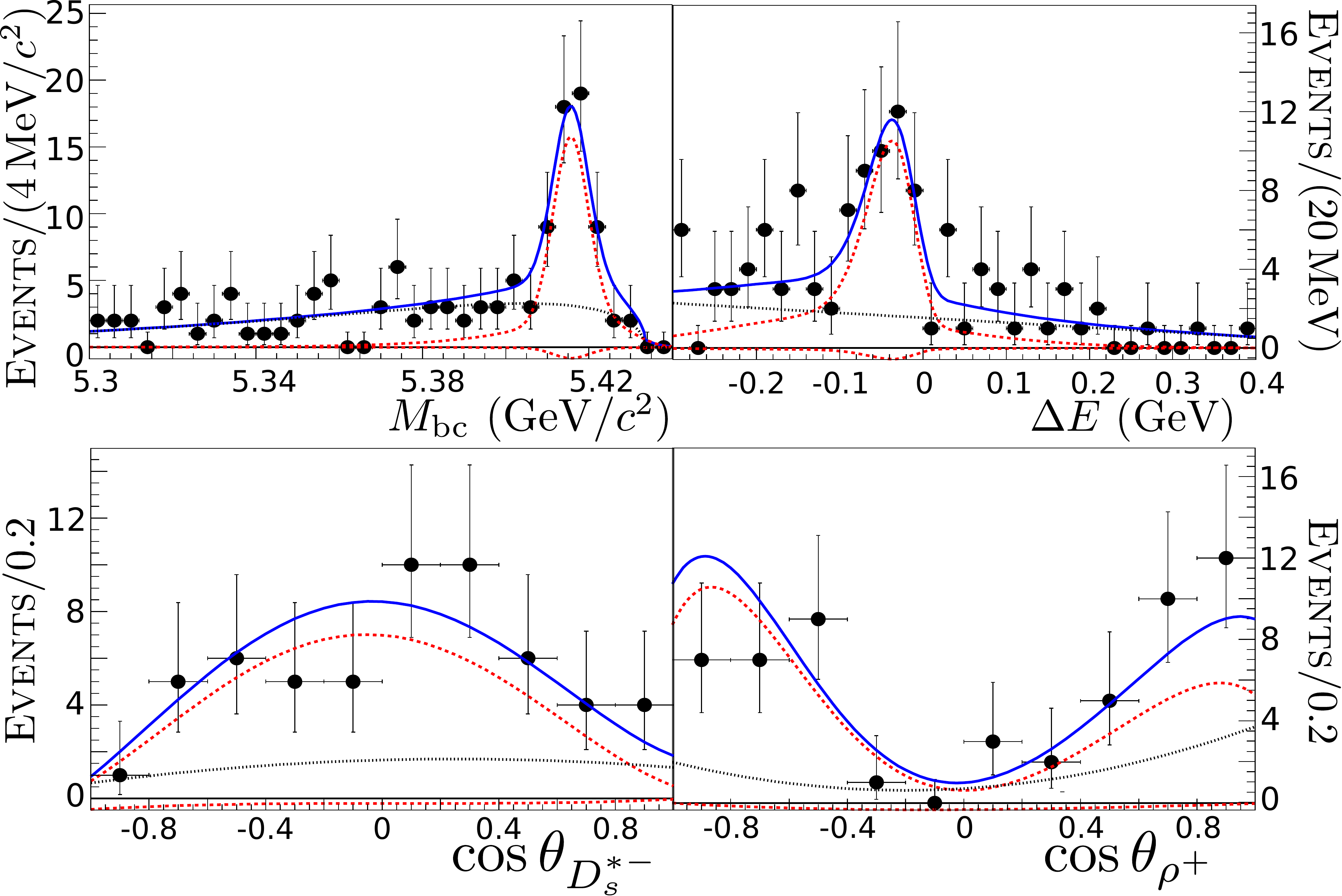}
	\caption{\label{fig3}Distributions for the $\bsdsstrho$ candidates.
          Top: $\mbc$ and $\deltae$ distributions, as in Fig.~\ref{fig2}.
	  Bottom: helicity distributions of the $D_s^{\ast -}$ (left) and $\rho^+$ (right)
	  with $\mbc$ and $\deltae$ restricted to the $\bsst\barbsst$ kinematic region.
	  The components of the total PDF (blue solid line) are shown separately:
	  the black-dotted curve is the background and
          the two red-dashed curve are the signal.
	  The large (small) signal shape corresponds to the longitudinal (transverse) component.}
      \end{figure}

      The fitted signal yields are listed in Table~\ref{tab:yields},
      while Figs.~\ref{fig2} and \ref{fig3} show the observed distributions
      in the $\bsst\barbsst$ signal region with the projections of the fit result.
      The significance is defined by $S=\sqrt{2\ln(\mathcal L_{\rm max}/\mathcal L_0)}$,
      where $\mathcal L_{\rm max}$ ($\mathcal L_0$) is the value at the maximum
      (with the corresponding yield set to zero)
      of the likelihood function convolved with a Gaussian distribution
      that represents the systematic errors of the yield.
      The linearity of the floating parameters in the region near the results
      has been extensively checked with MC simulations,
      as well as the statistical uncertainty of $f_L(\bsdsstrho)$,
      which lies near the limit of the physically allowed range $(0-1)$.
      
      \begin{table}
	\centering
	\caption{\label{tab:yields}Total efficiencies ($\varepsilon$),
	  signal yields ($N_S$) with statistical errors, 
	  and significance ($S$) including systematic uncertainties,
          for the three measured modes.}
	\begin{ruledtabular}
	  \renewcommand{\arraystretch}{1.3}
	  \begin{tabular}
	    {@{\hspace{0.2cm}}l@{\hspace{0cm}}c@{\hspace{0cm}}c@{\hspace{0cm}}c@{\hspace{0cm}}c@{\hspace{0.2cm}}}
	    \multicolumn{1}{c}{Mode}      &     Prod.~mode       & $\varepsilon$ (\%) & $N_S$                  &  $S$\\
	    \hline\hline
	    \multirow{3}{*}{$\bsdsstpi$}  & $\bsst\barbsst$  & $9.13$        & $53.4^{+10.3}_{-9.4}$  & $7.1\sigma$\\
	    & $\bsst\bar\bs+\bs\barbsst$   & --            & $-1.9^{+4.0}_{-2.9}$   & --\\
	    & $\bs\bar\bs$     & --            & $2.9^{+3.9}_{-3.0}$    & --\\
	    \hline
	    \multirow{3}{*}{$\bsdsrho$}   & $\bsst\barbsst$  & $4.40$        & $92.2^{+14.2}_{-13.2}$ & $8.2\sigma$\\
            & $\bsst\bar\bs+\bs\barbsst$   & --            & $-4.0^{+5.2}_{-3.7}$   & --\\
	    & $\bs\bar\bs$     & --            & $-3.0^{+5.7}_{-4.0}$   & --\\
	    \hline
	    $\bsdsstrho$                  & $\bsst\barbsst$  & --            & $77.7^{+14.6}_{-13.3}$ & $7.4\sigma$\\
	    \multicolumn{2}{@{\hspace{0.7cm}}l}{Longitudinal component}  & $2.66$        & $81.3^{+16.0}_{-14.9}$ & --\\
	    \multicolumn{2}{@{\hspace{0.7cm}}l}{Transverse component}           & $2.68$        & $-3.5^{+8.0}_{-6.1}$   &--\\
	  \end{tabular}
	\end{ruledtabular}
      \end{table}
      
      The dominance of the $\FiveS\to\bsst\barbsst$ mode is confirmed.
      For better precision,
      we therefore extract the branching fractions (BF)
      using only the yields in this mode. 
      Table~\ref{tab:result} shows the values obtained with the relations
      $\BR=N_S/(N_{\bs}\times\varepsilon)$, for the $\bsdsstpi$ and $\bsdsrho$ modes.
      The values for $\BR(\bsdsstrho)$ and $f_L=\fl$ are obtained by floating these two parameters in a fit where
      the longitudinal (transverse) yield is replaced by the relation $N_{\bs}\times \BR\times f_L\times \varepsilon_L$ ($N_{\bs}\times \BR\times (1-f_L)\times \varepsilon_T$),
      with $N_{\bs}$, $\varepsilon_T$ and $\varepsilon_L$ being fixed.
      Since the transverse yield fluctuated to a negative central value, $f_L>1$.
      The common systematic uncertainties on the BF are due to the errors on
      the integrated luminosity (1.3\%), $\sigmabb$ (4.6\%), 
      $f_s$ (15.0\%), $f_{\bsst\bar\bsst}$ (4.3\%), the $\ds$ BF (6.4\%), the $R_2$ cut (2.0\%),
      the tracking efficiency (4.0\%) and the charged-particle identification (5.4\%).
      In addition, uncertainties due to the MC statistics (1.6\%, 2.3\%, 1.5\%),
      the neutral-particle identification (8.8\%, 5.4\%, 8.8\%)
      and the PDF shapes (4.6\%, 4.7\%, 4.3\%)
      depend on the ($\bsdsstpi$, $\bsdsrho$, $\bsdsstrho$) mode. 
      The systematic errors on $f_L$ are due to the uncertainties in PDF shapes.

      \begin{table}
	\centering
	\caption{\label{tab:result}Top: measured BF values with statistical,
	  systematic (without $f_s$) and $f_s$ uncertainties,
	  and HQET predictions from the factorization hypothesis \cite{PLB_318_549}.
	  Bottom: BF ratios where several systematic uncertainties cancel out.
	  We use our previous measurement of $\BR(\bsdspi)$ \cite{PRL_102_021801}.}
	\begin{ruledtabular}
	  \renewcommand{\arraystretch}{1.3}
	  \begin{tabular}
	    {@{\hspace{0.3cm}}l@{\hspace{0.2cm}}c@{\hspace{0.2cm}}c@{\hspace{0.3cm}}c@{\hspace{0.3cm}}}
	    \multicolumn{1}{c}{Mode}      &     $\BR$ ($10^{-3}$)& HQET ($10^{-3}$)\\
	    \hline
	    $\bsdsstpi$&$2.4^{+0.5}_{-0.4}\pm0.3\pm0.4$&$2.8$\\
	    $\bsdsrho$&$8.5^{+1.3}_{-1.2}\pm1.1\pm1.3$&$7.5$\\
	    $\bsdsstrho$&$11.8^{+2.2}_{-2.0}\pm1.7\pm1.8$&$8.9$\\
	    \hline\hline
	    \multicolumn{1}{c}{Ratios}\\
	    \hline
	    \multicolumn{3}{@{\hspace{0.3cm}}l@{\hspace{0.5cm}}}{$\BR(\bsdsstpi)/\BR(\bsdspi)=0.65^{+0.15}_{-0.13}\pm0.07$}\\
	    \multicolumn{3}{@{\hspace{0.3cm}}l@{\hspace{0.5cm}}}{$\BR(\bsdsrho)/\BR(\bsdspi)=2.3\pm0.4\pm0.2$}\\
	    \multicolumn{3}{@{\hspace{0.3cm}}l@{\hspace{0.5cm}}}{$\BR(\bsdsstrho)/\BR(\bsdspi)=3.2\pm0.6\pm0.3$}\\
	    \multicolumn{3}{@{\hspace{0.3cm}}l@{\hspace{0.5cm}}}{$\BR(\bsdsstrho)/\BR(\bsdsrho)=1.4\pm0.3\pm0.1$}\\
	  \end{tabular}
	\end{ruledtabular}
      \end{table}
            
      Our values for the BF are in good agreement with predictions based on HQET and
      the factorization approximation \cite{PLB_318_549}.
      The large value of $f_L(\bsdsstrho)$ is consistent with
      the value measured for $B^0\to D^{\ast-}\rho$ decays \cite{PRD_67_112002}
      and with the predictions of Refs.~\cite{PRD_42_3732,ZPC_1_269}.
      
      In summary, we report the first observation of
      three CKM-favored exclusive $\bs$ decay modes, 
      we extract their branching fractions, and, 
      for $\bsdsstrho$, we measure the longitudinal
      polarization fraction.
      Our results are consistent 
      with theoretical predictions based on HQET \cite{PLB_318_549} and 
      are similar to analogous $B^0$ decay branching 
      fractions.
      The dominance of the unexpectedly large
      $\FiveS\to\bsst\barbsst$ mode \cite{PRL_102_021801} is confirmed.      

      \begin{acknowledgments}
	We thank the KEKB group for excellent operation of the
	accelerator, the KEK cryogenics group for efficient solenoid
	operations, and the KEK computer group and
	the NII for valuable computing and SINET3 network support.  
	We acknowledge support from MEXT, JSPS and Nagoya's TLPRC (Japan);
	ARC and DIISR (Australia); NSFC (China); MSMT (Czechia);
	DST (India); MEST, NRF, NSDC of KISTI, and WCU (Korea); MNiSW (Poland); 
	MES and RFAAE (Russia); ARRS (Slovenia); SNSF (Switzerland); 
	NSC and MOE (Taiwan); and DOE (USA).
      \end{acknowledgments}
      
      %Bibliography from BibTeX
      

\begin{thebibliography}{28}
\bibitem[{cha()}]{chargeconj}
	\bibinfo{note}{{Unless specified otherwise, charge-conjugated modes are implied
	    throughout.}}
\bibitem[{nov()}]{novofunc}
	\bibinfo{note}{{The Novosibirsk function is defined as
	    $f(x)=\exp[-\frac12(\ln^2(1+\Lambda(x-x_0))/\tau^2+\tau^2)]$ with $\Lambda=$
	    $\sinh(\tau\sqrt{\ln 4})/(\sigma\sqrt{\ln4})$. The parameters represent the
	    mean ($x_0$), the width ($\sigma$) and the tail asymmetry ($\tau$).}}
\end{thebibliography}
\end{document}